Running title:

Title:

# Visualizing septins in early Drosophila embryos

**Manos Mavrakis**

Aix Marseille Université, CNRS, Centrale Marseille, Institut Fresnel, UMR 7249, Faculté des Sciences Saint-Jérôme, 13013 Marseille, France

corresponding author: e-mail address: manos.mavrakis@univ-amu.fr

**Abstract**

Functional studies in Drosophila have been key for establishing a role for the septin family of proteins in animal cell division and thus extending for the first time observations from the budding yeast to animal cells. Visualizing the distribution of specific septins in different Drosophila tissues and, in particular, in the Drosophila embryo, together with biochemical and mutant phenotype data, has contributed important advances to our understanding of animal septin biology, suggesting roles in processes other than in cytokinesis. Septin localization using immunofluorescence assays has been possible due to the generation of antibodies against different

Drosophila septins. The recent availability of lines expressing fluorescent protein fusions of specific septins further promises to facilitate studies on septin dynamics. Here, we provide protocols for preparing early Drosophila embryos to visualize septins using immunofluorescence assays and live fluorescence microscopy. The genetic tractability of the Drosophila embryo together with its amenability to high resolution fluorescence microscopy promise to provide novel insights into animal septin structure and function.

**INTRODUCTION**

Studies in Drosophila have been instrumental in the history and development of the septin field since Drosophila is the first animal model system which established that the septin family of proteins existed and had significant roles in animals and not only in budding yeast (Fares, Peifer, & Pringle, 1995; Neufeld & Rubin, 1994). Flescher and coworkers had noted budding yeast Cdc10-like sequences in Drosophila and mouse (Flescher, Madden, & Snyder, 1993) and several other groups had come across mammalian septin genes in the early 1990s (Kumar, Tomooka, & Noda, 1992; Nakatsuru, Sudo, & Nakamura, 1994), but it was the identification and functional analysis of Pnut and DSep1 in Drosophila in the seminal studies by Neufeld and Rubin (Neufeld & Rubin, 1994) and Fares and colleagues (Fares et al., 1995) that laid down the groundwork for animal septin studies. Immunofluorescence studies using antibodies against Pnut and DSep1 in wild-type and *pnut* mutant tissues together with biochemical data provided key observations showing among others (1) that animal septins were also required for cytokinesis (Neufeld & Rubin, 1994), (2) a strong enrichment of septins in non-dividing cells, notably in the nervous system (Fares et al., 1995; Neufeld & Rubin, 1994), suggesting septin roles in processes other than in cytokinesis, and (3) that septins existed in heteromeric complexes (Fares et al., 1995). These studies were followed by the breakthrough work of Field and colleagues (C. M. Field et al., 1996), which isolated for the first time a native three-septin complex from Drosophila embryos that was able to polymerize into filaments in vitro, thus providing the first evidence for the polymerizing activity of septins, which had only been hypothesized at the time on the basis of electron microscopy studies of the budding yeast neck (Byers & Goetsch, 1976).

Follow-up studies in the Drosophila embryo advanced significantly our understanding of animal septin function. Using immunofluorescence assays in embryos that were maternally deprived of the septin Pnut, Adam and colleagues (Adam, Pringle, & Peifer, 2000) showed that (1) not all septins are required for cell divisions, (2) that not all septins function as part of a unique septin complex, and (3) that septins contribute to actin organization and are essential for normal embryonic development. Field and colleagues subsequently identified DAnillin as a septin-interacting partner that is required for recruiting septins to the tips of invaginating membranes during Drosophila embryo cellularization as well as to cytokinetic furrows in dividing embryonic cells (C. M. Field, Coughlin, Doberstein, Marty, & Sullivan, 2005). The DAnillin-septin interaction in Drosophila syncytial embryos was later shown to be regulated by the GTPase Ran (Silverman-Gavrila, Hales, & Wilde, 2008) and to be important for the ingression of pseudocleavage furrows whose tips are enriched in septins (Fares et al., 1995; Mavrakis, Rikhy, & Lippincott-Schwartz, 2009). Peanut distribution at those pseudocleavage furrows was also shown to depend on Diaphanous (Afshar, Stuart, & Wasserman, 2000) and Dynamin (Rikhy, Mavrakis, & Lippincott-Schwartz, 2015).

Su and colleagues (Su, Chow, Boulianne, & Wilde, 2013) later showed that septins localize not only to the tips of invaginating membranes during cellularization, but also to tubular extensions emanating from those tips, raising the tantalizing hypothesis that septin-mediated membrane tubulation contributes to local membrane remodeling. Biochemical work from this same study also suggested that heteromeric septin complexes other than the originally isolated Peanut-DSep1-DSep2 complex (for example, Peanut-DSep4-DSep5) might also be at work (Su et al., 2013). More recently, Drosophila septins were shown to cross-link actin filaments, and a direct septin-actin interaction was proposed to underly the organization of actin filaments in tight parallel bundles in the constricting rings at the tips of invaginating membranes in cellularizing embros (Mavrakis et al., 2014). This actin cross-linking activity of septins could also explain the reduced constriction rates and altered mechanical properties of cytokinetic rings in dividing embryonic epithelial cells mutant for septins (Guillot & Lecuit, 2013). The above studies highlight altogether how work in the Drosophila embryo has contributed to furthering our understanding of animal septin function.

Immunofluorescence assays have been central to all studies for providing snapshots of septin localization in wild-type and mutant conditions. The use of GFP-tagged septins opened up the study of septin dynamics in budding yeast (Cid, Adamikova, Sanchez, Molina, & Nombela, 2001; Lippincott, Shannon, Shou, Deshaies, & Li, 2001). However, the first Drosophila septin-GFP fusion (DSep2-GFP) was available only in 2008 (Christine M. Field, Maddox, Pringle, & Oegema, 2008; Silverman-Gavrila et al., 2008) with other fluorescent fusions only reported very recently (see Table 2). Thus, there is still a lot to learn about septin dynamics and their functional importance using fluorescent septin fusions in Drosophila. Although this chapter focuses on the use of the Drosophila embryo for investigating septin function, it is important to note that significant advances have also been contributed by septin studies in other Drosophila tissues (detailing of which is beyond the scope of this introduction).

The relative simplicity of the septin gene family in Drosophila (having 5 septin genes compared to 13 septin genes in human (Fung, Dai, & Trimble, 2014)) and the fact that the Drosophila embryo is amenable to high-resolution live fluorescence imaging makes it a powerful genetically tractable model system for studying a wide spectrum of septin functions in essential cell biological processes, including cell cycle regulation, cytoskeleton and membrane remodeling, and tissue morphogenesis. Here, we provide protocols for preparing Drosophila embryos for visualizing septins using immunofluorescense assays and live fluorescence microscopy.

## 1. Preparation of early Drosophila embryos for immunofluorescence detection of septins

We provide protocols for collecting, dechorionating, fixing and staining early Drosophila embryos with septin antibodies. We describe four fixation protocols that have been used for visualizing septins in the early Drosophila embryo (see Table 1). Each fixation protocol has different effects on the preservation of antigenicity and of sub-cellular or/and tissue structures (Muller, 2008). The choice of the protocol when co-staining embryos for septins and other cellular proteins will thus need to consider the specific requirements (epitope availability and preservation, subcellular localization) for all proteins in question. Slow formaldehyde fixation (section 1.1.1.) preserves well most cellular structures and epitopes and works for most antibodies. If the epitope is sensitive to methanol (which is otherwise used to physically remove the vitelline membrane), formaldehyde-fixed embryos can be devitellinized by hand (section 1.1.2.). For example, methanol fixation destroys the phalloidin-binding site on actin filaments, thus hand devitellinization is required for co-labeling septins and actin using fluorescent phalloidins (section 1.1.2.). Ice-cold methanol fixation (section 1.1.3.) has the advantage that almost all embryos are devitellinized, however it is a harsh treatment and destroys membranes. Alternatively, heat treatment followed by devitellinization in methanol (section 1.1.4.) might be required to preserve certain epitopes although the preservation of the structure is not very good (Miller, Field, & Alberts, 1989; Muller, 2008).

## 1.0. Embryo collection and dechorionation

**Materials and reagents**

- 60x15 mm petri dishes (10789241, Thermo Scientitic Nunc)

- small embryo collection cages that fit 60 mm petri dishes (59-100, Genesee Scientific) (Figure 1)

- apple juice agar plates for embryo collection

- baker's yeast paste

- egg basket with mesh for embryo collection (Figure 1)

- squirt bottle with distilled water

- squirt bottle with household bleach (4% sodium hypochlorite)

- paintbrush (11814, Ted Pella) for dispersing embryos

- dissecting stereomicroscope (e.g., Zeiss SteREO Discovery.V8)

1. Collect embryos of the genotype and developmental stage of interest on fresh yeasted plates following standard procedures (Cavey & Lecuit, 2008; Mavrakis, Rikhy, Lilly, & Lippincott-Schwartz, 2008).

2. Use a squirtbottle to add some water to the agar plate. With the help of the paintbrush bring the embryos and the yeast paste (also containing embryos) gently into suspension, and pour the embryo-yeast suspension into the egg basket. Wash with copious amounts of water using the squirt bottle to get rid of the yeast. Immerse the egg basket gently into a bottle lid containing 100% bleach and incubate for exactly 1 min while agitating gently to disperse the embryos. Wash the embryos immediately and very thoroughly with water using the squirt bottle to remove any residual bleach and pieces of chorion. Squirt water onto the sides of the basket to collect embryos onto the mesh. The bleach treatment removes the chorion, the outermost impermeable layer of the embryo.

*Notes:*

- Bleach loses its potency rapidly when exposed to air thus always pour fresh bleach and use it for dechorionation within 10 min. When embryos are dechorionated efficiently, their hydrophobic vitelline membranes stick to the sides of the basket and embryos also clump together. Inspect the embryos visually under the microscope (10-20x magnification) to ensure that dechorionation is complete. The micropyle at the anterior end of the vitelline membrane should be intact (Figure 1). If dorsal appendages or pieces of chorion are still observed, place embryos back to the bleach for 5 s, rinse thoroughly with water and check again. Do not overexpose embryos to bleach or they will be damaged.

- You need to prepare scintillation vials containing the appropriate fixative solutions (section 1.1.) right before dechorionating embryos. If needed, you can place the egg basket on a clean lid of a petri dish and add water to keep the embryos hydrated.

3. Proceed immediately with embryo fixation.

[Insert Figure 1 here]

**Figure 1.** Drosophila embryo collection and dechorionation. Image in A shows an embryo collection cage with adult flies over a yeasted apple juice agar plate. Image in B shows an egg basket made from a powder funnel glued to a 100-μm mesh for collecting and dechorionating embryos. Image in C shows a dechorionated Drosophila embryo imaged with transmitted light (A, P, D, V pointing to the anterior, posterior, dorsal and ventral axes of the embryo, respectively). The arrowhead points to the micropyle at the anterior end of the vitelline membrane.

## 1.1. Embryo fixation

**Materials and reagents**

- glass scintillation vials (986546, Wheaton Science Products)

- fine paintbrush (11810, Ted Pella) for transferring embryos

- heptane (anhydrous, 99%) (246654, Sigma)

- 37% w/w formaldehyde solution (252549, Sigma)

- methanol (anhydrous) (322415, Sigma)

- phosphate-buffered saline (PBS) solution

- Tween 20 (P1379, Sigma)

- 30% bovine serum albumin (BSA) solution (A7284, Sigma)

- Triton X-100 (X100, Sigma)

- sodium chloride (S9888, Sigma)

- Pasteur pipettes
- apple juice agar plates (see section 1.0)
- egg basket (see section 1.0)
- scalpel or razor blade for cutting out pieces of agar
- sewing needle for handling embryos on agar
- Moria nickel plated pin holder (26016-12, Fince Science Tools) for holding sewing needles
- hypodermic needles for hand devitellinization (BD Microlance 3 - 30G x 1/2" (0.3mm x 13mm needles) (304000, BD)
- 12-mm wide double-stick Scotch tape for hand devitellinization (3M)
- 1.5 mL microcentrifuge tubes
- aluminium foil
- orbital shaker for fixing embryos (Rotamax 120, Heidolph)
- test tube rocker for staining embryos (Vari-Mix or Speci-Mix Test Tube Rocker, Thermo Scientific)
- dissecting stereomicroscope (e.g., Zeiss SteREO Discovery.V8)
- primary and secondary antibodies (see Table 1 for septin antibodies)
- Aqua-Poly/Mount medium for mounting embryos (18606-20, Polysciences)
- glass slides 76 x 26 mm (10090431, Thermo Scientific Menzel)
- glass coverslips 22 x 50 mm for mounting embryos (100266, Dutscher)
- glass coverslips 24 x 32 mm for hand devitellinization (100035, Dutscher)

### 1.1.1. Slow formaldehyde fixation and methanol devitellinization

1. Prepare a scintillation vial with 5 mL of **heptane** and 5 mL of **3.7% formaldehyde** in PBS. Blot the egg basket gently on paper tissue to remove excess water. Using a slightly wet paintbrush and working under the microscope, transfer the dechorionated embryos to the heptane (upper) phase of the scintillation vial. Embryos will float at the interface between the heptane and the aqueous fixative phases. Cap and label the vial and swirl on an orbital shaker (100-150 rpm) for 20-30 min.

*Notes:*

- Heptane serves to permeabilize the vitelline membrane so that formaldehyde penetrates and fixes the embryo.

- After transfering embryos to the heptane phase, inspect briefly the embryos at the interface of the two phases under the microscope. The stages of development are clearly visible thus you can rapidly assess the enrichment of the collected embryos regarding the embryonic stage of interest before proceeding to fixation and staining.

2. Under the hood use a Pasteur pipette to draw off completely the aqueous fixative (lower) phase, taking care to leave behind the embryos at the bottom of the heptane phase.

3. Add immediately 5 mL of **methanol** and vortex for 30 s. Devitellinized embryos will sink to the bottom of the tube. Using a new Pasteur pipette remove carefully the heptane and methanol phases together with the non-devitellinized embryos at their interface that did not sink, taking care not to aspirate the dechorionated embryos at the bottom of the vial.

4. Add immediately 5 mL of fresh methanol and allow embryos to sink to the bottom. Use a new Pasteur pipette to remove the methanol leaving the embryos behind and add fresh methanol. Repeat twice with fresh methanol to remove residual heptane and formaldehyde.

*Note:* You can store embryos in methanol at 4°C or at -20°C at this point.

5. Rinse a new Pasteur pipette with methanol (to avoid that embryos stick to its sides) and use it to transfer the embryos from the scintillation vial to a microcentrifuge tube. Allow embryos to sink to the bottom of the tube, aspirate carefully the methanol and replace with fresh **PBST (PBS with 0.1% v/v Tween 20) containing 0.1% BSA**. Allow embryos to settle to the bottom of the tube and repeat twice to remove residual methanol. Rehydrate embryos by washing them three times in fresh PBST containing 0.1% BSA for 5 min each on a test tube rocker.

6. Proceed with embryo staining (section 1.2.). Figure 2 (A and C) shows embryo stainings using this protocol.

### 1.1.2. Slow formaldehyde fixation and hand devitellinization

1. Follow the instructions of step 1 of section 1.1.1. to fix embryos at the interface of 5 mL of **heptane** and 5 mL of **7.4% formaldehyde** in PBS for 20-30 min.

*Note:* The higher formaldehyde concentration helps devitellinize embryos more easily.

2. Use a Pasteur pipette to remove carefully both the heptane and the formaldehyde phases, taking care not to aspirate the embryos. Use a a 200-µL pipette tip to remove as much heptane/formaldehyde as possible. Fill immediately the scintillation vial with **PBST containing 0.1% BSA** and shake vigorously to disperse the embryos, which will sink to the bottom of the vial.

3. Prepare a sticky surface for hand-devitellinizing the embryos. Cover the surface of a clean 24 x 32 mm coverslip with two stripes of double-stick tape, letting 0.5 cm of tape extend off each side of the coverslip.

4. Rinse a new Pasteur pipette with PBST containing 0.1% BSA (to avoid that embryos stick to its sides) and transfer embryos from the scintillation vial to an egg basket.

*Note:* Place the basket with the embryos on the lid of a petri dish and add **PBST containing 0.1% BSA** to keep embryos hydrated while transfering them to the agar.

5. Use a scalpel to cut out a piece of agar ($\approx$ 2.5 x 3.5 cm, roughly the same size as the tape-coated coverslip) from an apple juice agar plate and place the agar on a clean slide. Blot gently the egg basket on paper tissue and use a slightly wet paintbrush to transfer the embryos from the basket to the agar under the microscope.

*Note:* If there are too many embryos and more time is needed, add a few drops of PBST containing 0.1% BSA to the embryos on the agar to keep them hydrated until all embryos have been transfered to the agar.

6. Under the microscope use a sewing needle to disperse the embryos in the PBST solution on the agar so that they are not clumped. Wait for most of the solution to be absorbed by the agar without letting embryos dry out. Invert immediately the coverslip covered by double-stick tape (step 3) onto the embryos. Press gently to stick all the embryos onto it, place it immediately onto a clean lid of a petri dish and cover the embryos with **PBST containing 0.1% BSA**.

*Note:* Use the short tape extension to stick the coverslip onto the lid.

7. Remove the vitelline membrane under the microscope using a 30-gauge hypodermic needle. To achieve this, poke a small hole in the vitelline membrane at one end of the embryo and gently push the embryo out through the hole by pushing from the opposite end. The vitelline membrane will remain stuck to the tape and the embryo will float in solution. The embryonic stages of development are clearly visible. Rinse a 200-μL pipette tip with PBST containing 0.1% BSA (to avoid that embryos stick to its sides) and transfer hand-devitellinized embryos of the stage of interest into a microcentrifuge tube. Once all embryos have been collected, proceed with embryo staining (section 1.2.). Figure 2B shows an embryo staining using this protocol.

### 1.1.3. Methanol fixation

1. Prepare a scintillation vial with 5 mL of **heptane** and 5 mL of **ice-cold methanol**. Blot the egg basket gently on paper tissue to remove excess water. Using a slightly wet paintbrush and working under the microscope, transfer the dechorionated embryos to the heptane (upper) phase of the scintillation vial. Embryos will float at the interface between the heptane and the methanol phases. Let embryos at the interface for 30 s, then vortex for 30 s. Devitellinized embryos will sink to the bottom of the tube. Using a Pasteur pipette remove carefully the heptane and methanol phases together with the non-devitellinized embryos at their interface that did not sink, taking care not to aspirate the dechorionated embryos at the bottom of the vial.

2. Use a new Pasteur pipette to remove the methanol leaving the embryos behind and add fresh methanol. Repeat twice with fresh methanol to remove residual heptane.

3. Follow the instructions of step 5 of section 1.1.1. to rehydrate the embryos in PBST containing 0.1% BSA and then proceed with embryo staining (section 1.2.).

### 1.1.4. Heat-methanol fixation

1. Prepare a scintillation vial with 5 mL of **heat-fix solution (0.03% v/v Tritox X-100, 68 mM NaCl)** and close the lid loosely. Microwave the solution on high power for a few seconds and stop right before it starts boiling (the solution becomes slightly opaque).

2. Use a slightly wet paintbrush to transfer dechorionated embryos to the **hot heat-fix solution** and swirl gently to disperse the embryos. Place the vial immediately on ice and add **ice-cold heat-fix solution** to fill the scintillation vial. Leave the vial on ice for 2 min.

3. Pour off the solution using a 200-µL pipette tip to remove as much solution as possible. Replace with 5 mL heptane and 5 mL methanol and proceed to devitellinization and embryo rehydration (steps 3-5 of section 1.1.1.).

*Note:* If the epitope is sensitive to methanol, you can replace the heat-fix solution with fresh PBST containing 0.1% BSA and proceed with hand devitellinization (steps 3-7 of section 1.1.2.).

[Insert Table 1 here]

**Table 1.** Drosophila septin antibodies used for immunofluorescence studies in the Drosophila embryo.

### 1.2. Embryo staining with antibodies

1. Incubate rehydrated embryos in **PBST containing 10% BSA** for at least 10 min on a test tube rocker to block non-specific epitopes that could otherwise bind to antibodies.

*Note:* You can store a stock of PBST containing 10% BSA at 4°C for several months.

2. Dilute primary antibodies in **PBST containing 0.1 % BSA**. See Table 1 for dilutions that have been used with septin antibodies. Allow embryos to settle to the bottom, remove the blocking solution and replace with 200 μL of the primary antibody solution. Incubate overnight at 4°C on a test tube rocker.

*Note:* Alternatively, you can incubate embryos with primary antibodies for 2 h at room temperature, but we strongly recommend an overnight incubation at 4°C to minimize non-specific binding of antibodies and thus reduce the background.

3. Rinse embryos with fresh **PBST containing 0.1 % BSA**, then wash three times in fresh PBST containing 0.1 % BSA for 10 min each time on a test tube rocker.

*Note:* Take care to collect embryos that might have stuck to the tube cap before rinsing and washing.

4. Dilute secondary antibodies in **PBST containing 0.1 % BSA**. Incubate embryos with with 200 μL of secondary antibodies for 2 h at room temperature on a test tube rocker. Cover the tubes with aluminum foil to protect the fluorophores from light.

5. Repeat step 3.

6. The embryos are ready to mount. Aspirate most of the wash solution leaving behind embryos in 100 μL. Rinse briefly a 200-μL pipette tip with PBST (to avoid that embryos stick to its sides), aspirate gently the embryos 2-3 times to disperse them and transfer immediately on a clean slide. Aspirate carefully as much PBST as possible and add immediately 2-3 drops of Aqua-Poly/Mount while dispersing the embryos. Drop gently a clean 22 x 50 mm coverslip onto the mounting medium and allow capillary forces to spread the mounting medium containing the embryos. Store the slide with the mounted embryos flat at 4°C in the dark for 24 h to allow Aqua-Poly/Mount to solidify before imaging.

[Insert Figure 2 here]

**Figure 2**. Examples of septin localization in the early Drosophila embryo using immunofluorescence assays with slow formaldehyde fixation and methanol (A, C) or hand (B) devitellinization. Images have been acquired with a laser scanning confocal microscope using a Zeiss Plan-Apo 63x/1.4 oil objective. (A,B) Sagittal sections of cellularizing embryos showing the localization of two septins, Pnut and DSep1, nonmuscle myosin heavy chain (Zip), the lateral membrane protein Scribble (Scb) and F-actin (infered by phalloidin). (C) End-on views of embryos at the end of cellularization stained for Pnut and for the polarity protein DPatj. Pnut and DPatj are found both at apicolateral membranes (top panel) and at the contractile actin-myosin rings at the membrane front (bottom panel).

## 2. Preparation of early Drosophila embryos for live fluorescence imaging of fluorescent protein fusions of septins

Here we provide a protocol where we immobilize dechorionated embryos on heptane-glue coated coverslips for live fluorescence imaging. We cover embryos with halocarbon oil, which prevents dehydration while allowing gas exchange, and image on an inverted microscope using oil or water immersion medium depending on the application. Alternatively, we cover embryos with PBS and use a water dipping objective for two-photon imaging.

**Materials and reagents**

- Halocarbon 200 Oil (25073, TEBU-BIO)

- heptane glue: Cut brown packaging tape (premium grade packaging tape Tesa 4124) in small pieces (5-10 cm long), roll each piece onto itself with the glue side facing out and place them in a 50-mL Falcon tube. Pack the tape pieces tightly to fill the tube. Fill the tube with heptane under a hood and let the tubes on a rolling or shaking platform for 2 days to extract the glue. Remove the tape pieces and centrifuge at 20,000 g for 1 h. Recover the supernatant, distribute in microcentrifuge tubes and centrifuge again at 20,000 g for 30 min. Repeat twice. The supernatant should be transparent and light yellowish in color. Distribute in glass scintillation vials, seal with parafilm and store at room temperature. If the glue is not sticky enough for embryos, let the vials open under the hood overnight to concentrate the glue. If autofluorescent glue spots are present on the coverslip and interfere with surface imaging, longer centrifugations are required.

- apple juice agar plates, fine paintbrush for transferring embryos on agar, Pasteur pipettes, sewing needles for aligning and orienting embryos on agar (see materials of section 1.1.)

- dissecting stereomicroscope (e.g., Zeiss SteREO Discovery.V8) for aligning embryos

- glass coverslips 24 x 32 mm for imaging (100035, Dutscher). Hold these coverslips in place on the stage with a slide holder or slide clamps.

- round coverslips 25 x 25 mm for imaging (140499, Dutscher). Hold these coverslips in place using a coverslip holder (Attofluor® Cell Chamber, 10604043, Fisher Scientific) and a round slide holder on the stage.

[Insert Table 2 here]

**Table 2.** Drosophila septin fluorescent protein fusions used in the Drosophila embryo.

## 2.0. Embryo preparation and mounting

1. Prepare a coverslip with heptane glue. Use a Pasteur pipette to place a line (for a rectangular coverslip) or drop (for a round coverslip) of heptane glue on the coverslip. Allow the heptane to evaporate for 1-2 min. The glue-coated coverslip is ready to use. Cover with the lid of a clean petri dish to protect it from dust.

*Note:* The choice of coverslip depends on the application. When using oil immersion (which allows the use of high NA oil objectives), rectangular coverslips that are held in place by a slide holder or clamps tend to be pulled by the viscous drag of the oil during z-acquisition (but this is not a problem when using water immersion). Round coverslips held in place by the coverslip holder do not present this problem. Also, round coverslips allow the rotation of the holder on the stage thus providing flexibility for (re)orienting embryos on the stage, if needed.

2. Collect and dechorionate embryos as described in steps 1-2 of section 1.0. Use a scalpel to cut out a piece of agar (≈ 3 x 3 cm) from an apple juice agar plate and place the agar on a clean slide. Blot gently the egg basket on paper tissue and use a slightly wet paintbrush to transfer the embryos from the basket to the agar under the microscope.

3. Use a sewing needle to align and orient the embryos on the agar so that the region of interest that will be imaged faces up. Invert the glue-coated coverslip from step 1 onto the embryos and press gently against the agar to glue all embryos. Cover immediately with halocarbon oil. The embryos are ready to be imaged. The whole procedure takes about 20 min thus collect embryos young enough so that they reach the developmental stage of interest when brought to the microscope for imaging.

*Notes:*

- If large numbers of embryos are to be images, several parallel lines of embryos can be prepared. In this case, the spacing between embryos should be one embryo width apart to prevent defects from anoxia.
- Use only as much halocarbon oil as needed to cover the embryos. If you use too much oil, it will spread during imaging and cause embryo dehydration.
- To be able to recognize specific stages of embryonic development and thus select specific embryos for mounting, practice after consulting free online resources for the Drosophila community (for example, The Interactive Fly, http://www.sdbonline.org/sites/fly/aimain/1aahome.htm) and reference textbooks on embryonic development (Campos-Ortega & Hartenstein, 1997).

## 2.1. Embryo imaging

We routinely image fluorescent septin fusions close to the embryo surface with a spinning disk microscope. However, light scattering leads to important fluorescence quenching with increasing imaging depth. Thus, for imaging septin fusions deeper (>15 μm away from the surface) in the embryo, for example at the constricting tips of invaginating membranes during cellularization, we routinely use two-photon microscopy (Figure 3). Considerations related to live fluorescence imaging in Drosophila embryos (choice of the microscope, of the objective and optimization of imaging conditions, as well as specificities and optimization regarding the different fluorescence-based techniques for studying protein dynamics) are beyond the scope of this chapter, but are detailed in the literature (Cavey & Lecuit, 2008; Mavrakis, Rikhy, Lilly, & Lippincott-Schwartz, 2008).

[Insert Figure 3 here]

**Figure 3**. Live imaging of fluorescent protein septin fusions. End-on (A) and sagittal (B) views of cellularizing embryos expressing mCherry-Pnut (A) or DSep2-GFP (B), imaged by spinning disk (A) or two-photon (B) microscopy. Two-photon microscopy allows us to visualize septins at the contractile ring throughout cellularization (~35 μm away from the surface at the end of cellularization) and to measure changes in septin levels during ring constriction.

**Figure legends**

**Figure 1**. Drosophila embryo collection and dechorionation. Image in A shows an embryo collection cage with adult flies over a yeasted apple juice agar plate. Image in B shows an egg basket made from a powder funnel glued to a 100-μm mesh for collecting and dechorionating embryos. Image in C shows a dechorionated Drosophila embryo imaged with transmitted light (A, P, D, V pointing to the anterior, posterior, dorsal and ventral axes of the embryo, respectively). The arrowhead points to the micropyle at the anterior end of the vitelline membrane.

**Figure 2**. Examples of septin localization in the early Drosophila embryo using immunofluorescence assays with slow formaldehyde fixation and methanol (A, C) or hand (B) devitellinization. Images have been acquired with a laser scanning confocal microscope. (A,B) Sagittal sections of cellularizing embryos showing the localization of two septins, Pnut and DSep1, nonmuscle myosin heavy chain (Zip), the lateral membrane protein Scribble (Scb) and F-actin (infered by phalloidin). (C) End-on views of embryos at the end of cellularization stained for Pnut and for the polarity protein DPatj. Pnut and DPatj are found both at apicolateral membranes (top panel) and at the contractile actin-myosin rings at the membrane front (bottom panel).

**Figure 3**. Live imaging of fluorescent protein septin fusions. End-on (A) and sagittal (B) views of cellularizing embryos expressing mCherry-Pnut (A) or DSep2-GFP, imaged by spinning disk (A) or two-photon (B) microscopy.

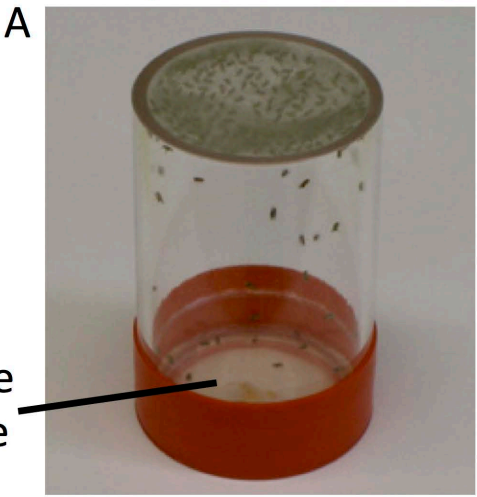
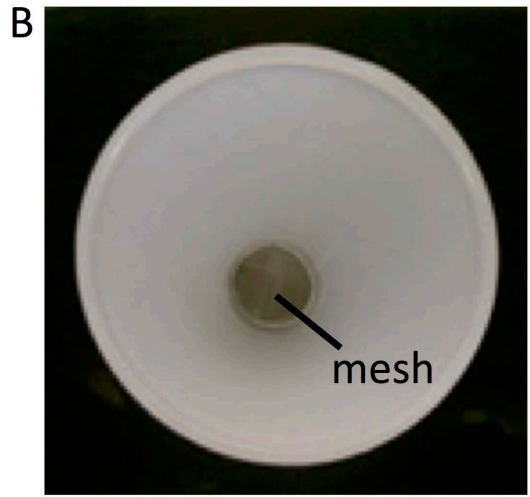
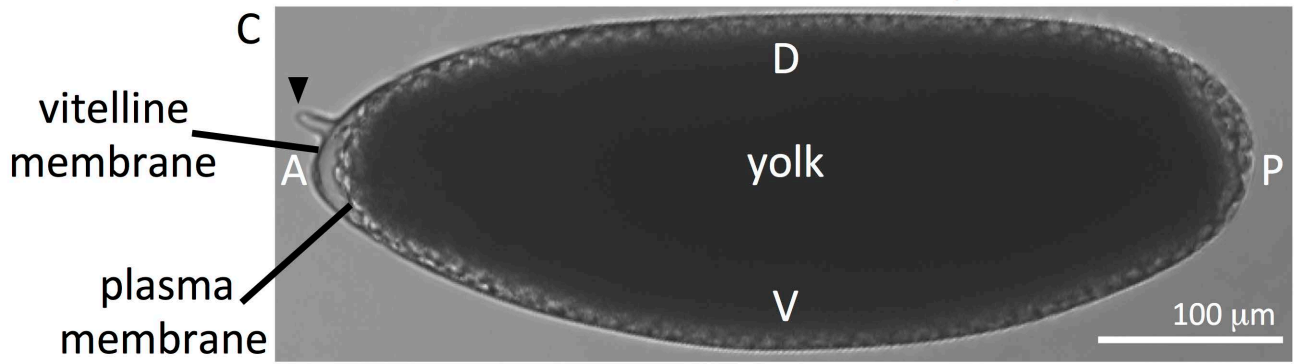

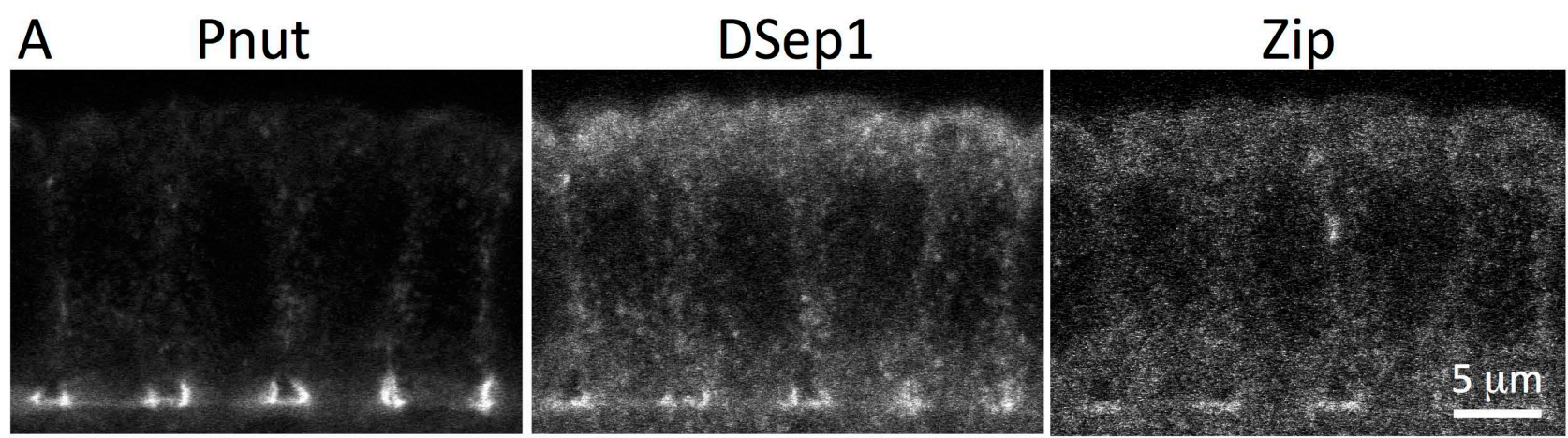
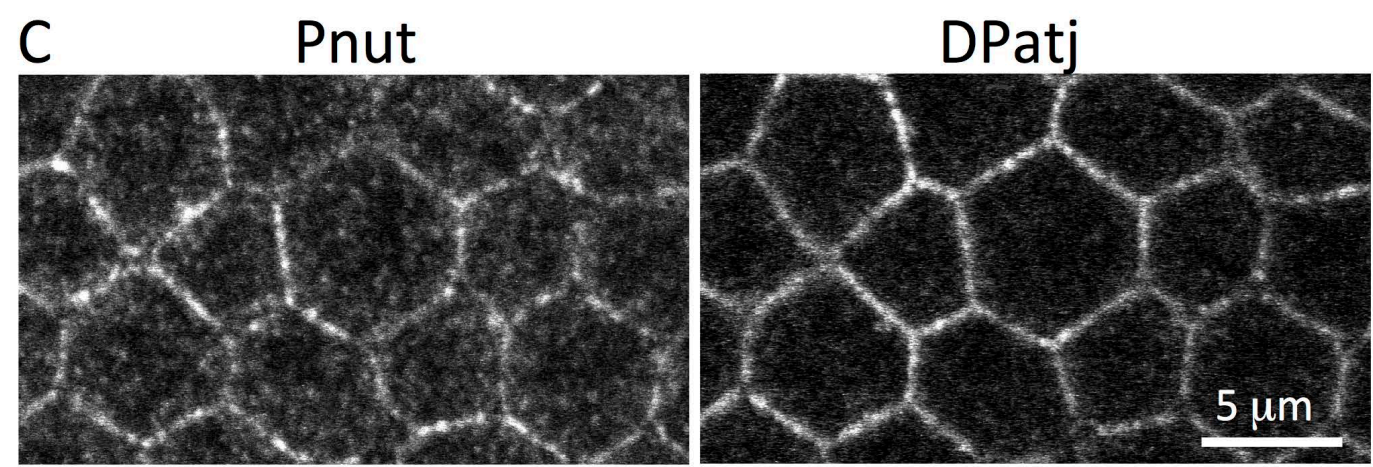
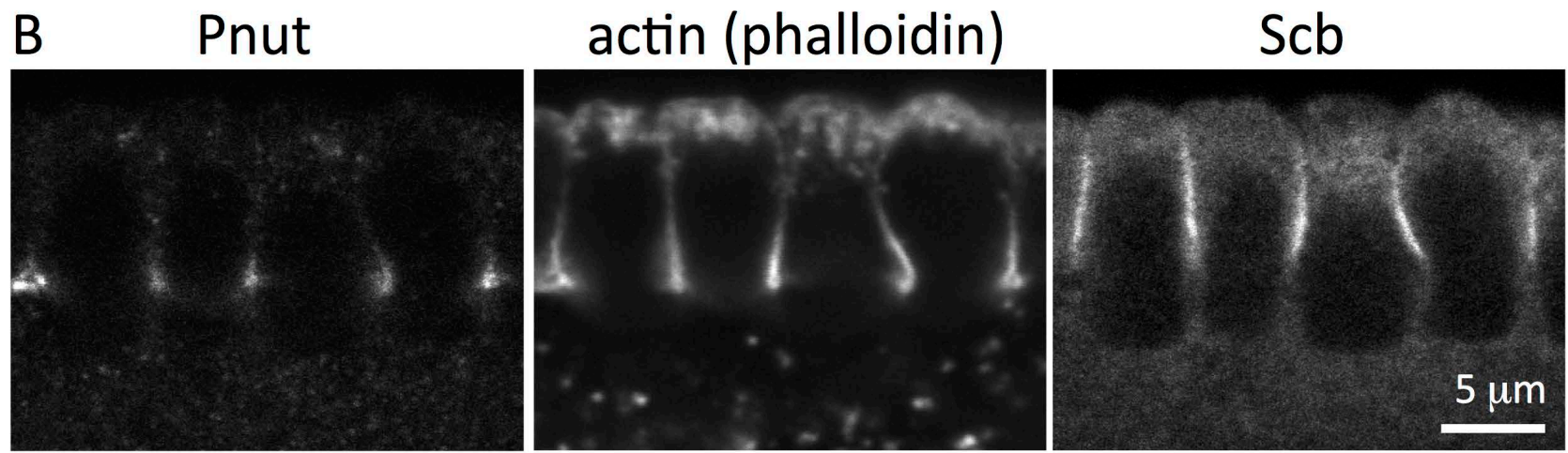
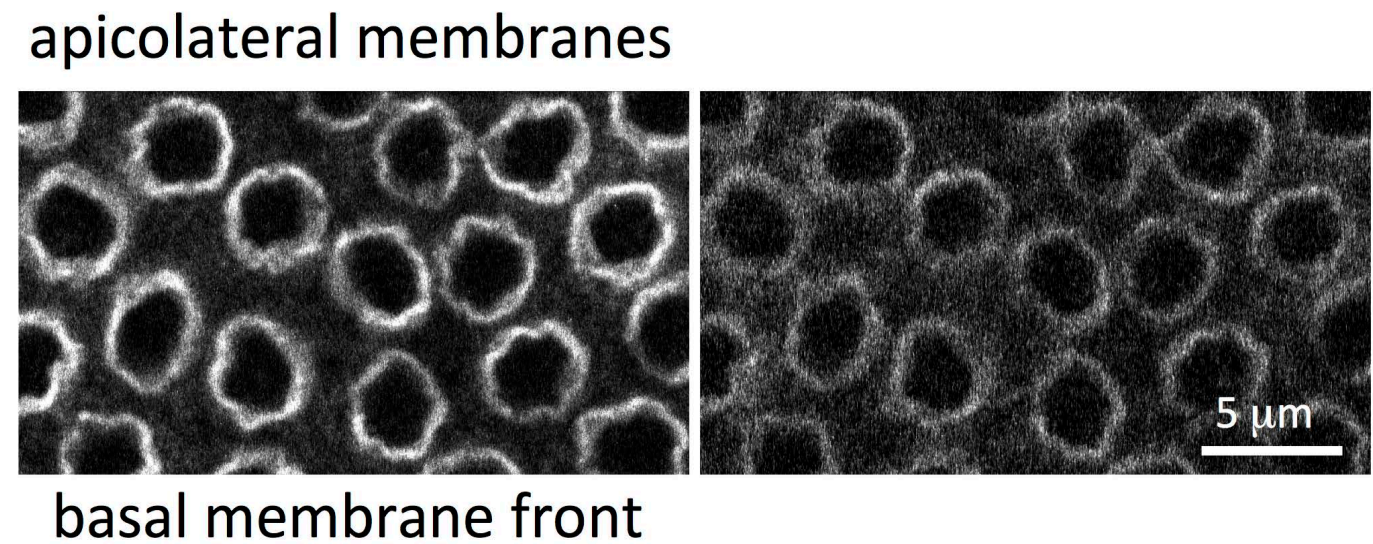

apicolateral membranes

basal membrane front

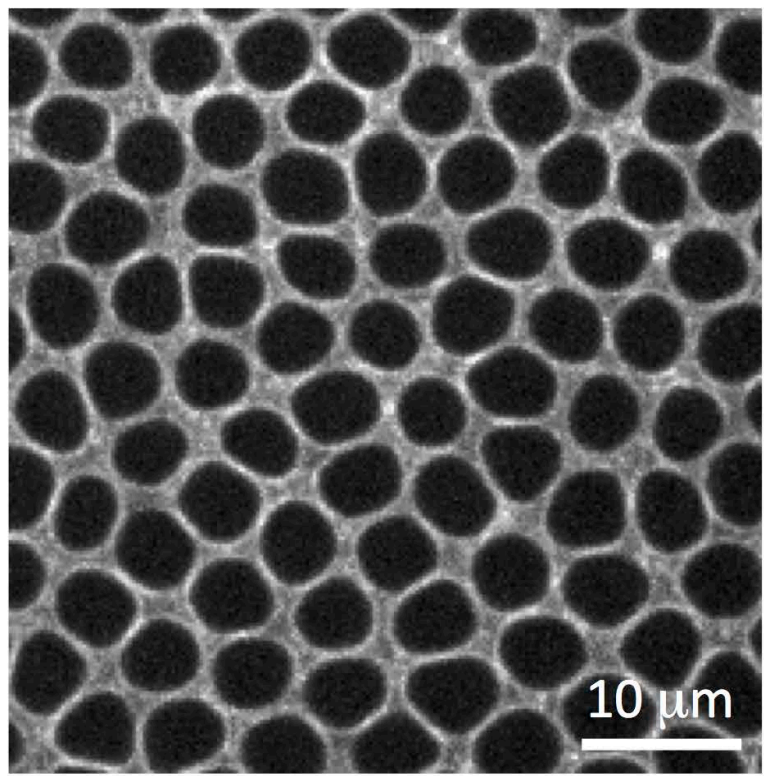 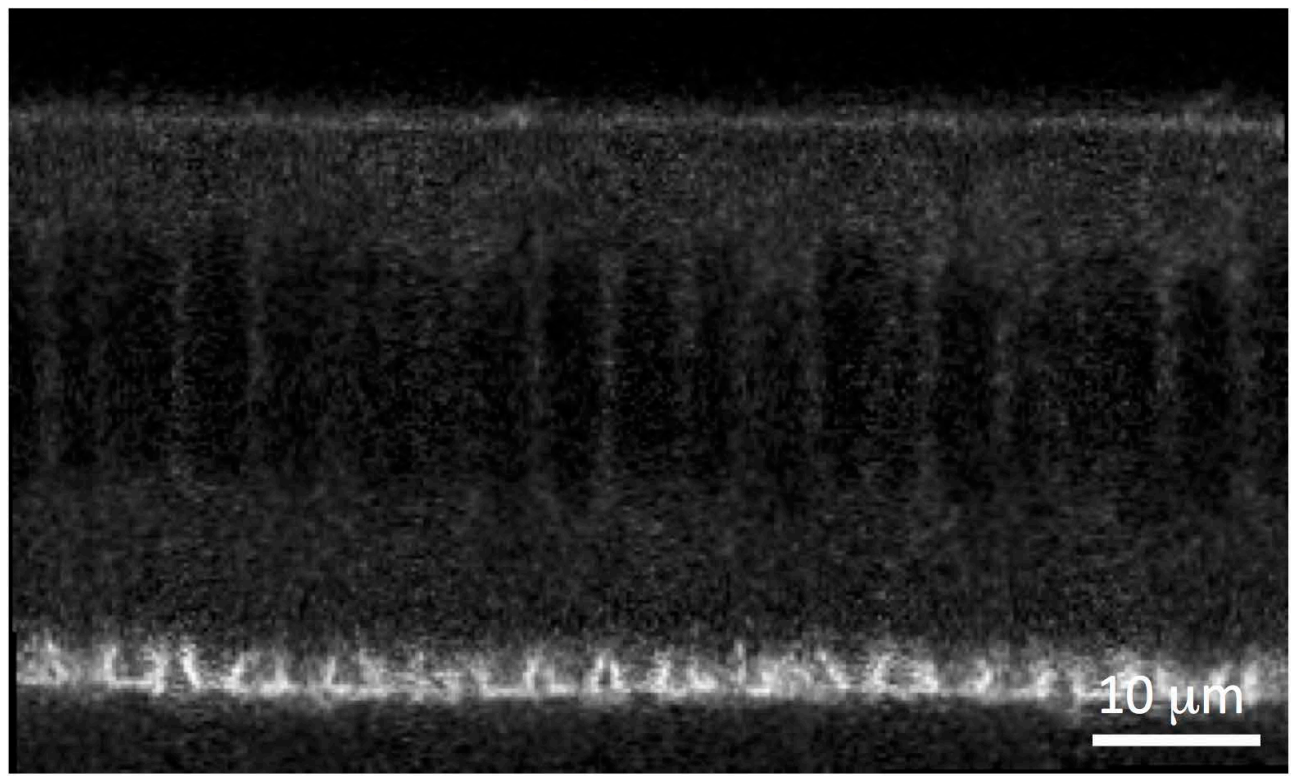

| A | mCherry-Pnut | B | DSep2-GFP |

spinning disk microscopy | two-photon excitation

# Table 1. Drosophila septin antibodies used for immunofluorescence studies in the Drosophila embryo*

| septin | antibody name | antigen | species | originally reported in | fixation | dilution for IF[a] |
|---|---|---|---|---|---|---|
| Pnut | 4C9 [b] | N-terminal 116 aa (GST fusion) | mouse | (Neufeld & Rubin, 1994) | heptane:4% paraformaldehyde 1:1 | 1:4 (Neufeld & Rubin, 1994); 1:3 (Adam, Pringle, & Peifer, 2000) |
| Pnut | KEKK | C-terminal 14 aa | rabbit | (C. M. Field et al., 1996) | cold methanol:heptane 1:1 (C. M. Field, Coughlin, Doberstein, Marty, & Sullivan, 2005) | 1 µg/mL (C. M. Field et al., 2005); 5 µg/mL (Adam et al., 2000) |
| DSep1 | | full-length DSep1 (TrpE-Sep1 and MalE-Sep1 fusions) | rabbit | (Fares, Peifer, & Pringle, 1995) | heptane:4% formaldehyde 1:1 | 1:50 (Adam et al., 2000) |
| DSep2 [c] | | C-terminal 327 aa (GST fusion) | rabbit | (C. M. Field et al., 1996) | heat-methanol fixation (Adam et al., 2000) | 1:20 (Adam et al., 2000) |
| DSep1 | Sep1-95 | N-terminal 15 aa | rat | (Mavrakis et al., 2014) | heptane:4% formaldehyde 1:1 | 1:250 |
| DSep2 [d] | Sep2-92 | 15 aa-peptide near the N-terminus | guinea pig | (Mavrakis et al., 2014) | heptane:4% formaldehyde 1:1 | 1:250 |

[a] Most of these antibodies have also been used for immunoblotting or immunoprecipitation experiments. Refer to the respective citations for the dilutions used for those applications.

[b] This antibody is available from the Developmental Studies Hybridoma Bank: http://dshb.biology.uiowa.edu/ (4C9H4 supernatant).

[c] These anti-Sep2 antibodies were found to cross-react weakly with DSep5 (Adam et al., 2000; Christine M. Field, Maddox, Pringle, & Oegema, 2008), reflecting the 73% amino acid sequence identity betwen DSep2 and DSep5. Also, the peptide sequence used to identify the 50 kDa band as DSep2 in immunopurified septin complexes (C. M. Field et al., 1996) was EMLIR (Chris Field, personal communication), which is also contained in the DSep5 sequence (the full sequence of which was only deposited in 1999) (Christine M. Field et al., 2008). Thus, it is possible that the detected signal in immunofluorescence and immunoblotting studies using this antibody (also) corresponds to DSep5.

[d] Although these DSep2 antibodies specifically recognize DSep2 in Western blots on purified DSep1-DSep2-Pnut complexes (Mavrakis et al., 2014), they cross-react with additional proteins in the early Drosophila embryo both in Western blots and by immunofluorescence (Mavrakis et al., 2014).

* New rabbit anti-Pnut (using full-length His-tagged Pnut as an antigen) (Huijbregts, Svitin, Stinnett, Renfrow, & Chesnokov, 2009), rabbit anti-DSep1 (using full-length His-tagged DSep1 as an antigen) and rabbit anti-DSep2 (using full-length DSep2 as an antigen) (Akhmetova, Balasov, Huijbregts, & Chesnokov, 2015) antibodies have been recently generated and used for biochemical studies but their usability for embryo immunofluorescence assays has not been tested.

## Table 2. Drosophila septin fluorescent protein fusions used in the Drosophila embryo*

| septin fusion[a] | Drosophila stock number[b] | promoter for expression | originally reported in |
|---|---|---|---|
| DSep2-EGFP | 26257 | *sep2* (genomic transgene) | (Field, Maddox, Pringle, & Oegema, 2008; Silverman-Gavrila, Hales, & Wilde, 2008) |
| DSep1-EGFP | 51346 | UASp | (Su, Chow, Boulianne, & Wilde, 2013) |
| DSep4-EGFP | 51345 | UASp | (Su et al., 2013) |
| DSep5-EGFP | 51344 | UASp | (Su et al., 2013) |
| DSep5-mRFP | 56492 | UASp | FlyBase (http://flybase.org/) by Andrew Wilde (U. Toronto) |
| Pnut-EGFP |  | UASp | (Su et al., 2013) |
| mCherry-Pnut |  | UASp | (Guillot & Lecuit, 2013) |

[a] Apart from the mCherry-Pnut transgene, where mCherry is fused to the N-terminus of Pnut, the other septin transgenes encode septins whose C-termini are tagged with GFP.

[b] These stocks are available from the Bloomington Drosophila Stock Center (http://flystocks.bio.indiana.edu/).

* A Venus (variant of yellow fluorescent protein) DSep1 CPTI trap line (DGRC#115423) was recently generated and is available through the Kyoto Drosophila Genetics Resource Center (https://kyotofly.kit.jp/cgi-bin/stocks/search_res_list.cgi?DB_NUM=1&PREDEF=ProteinTrapCamb). Venus is fused to the N-terminus of DSep1 in this line.